\documentclass[%
 aip,
 amsmath,amssymb,
 reprint,%
]{revtex4-1}

\usepackage{graphics}
\usepackage{amsmath}
\usepackage{amsfonts}
\usepackage{latexsym}
\usepackage{amsbsy}
\usepackage{epstopdf}
\usepackage{graphicx}
\usepackage{dcolumn}
\usepackage{bm}

\usepackage[utf8]{inputenc}
\usepackage[T1]{fontenc}
\usepackage{mathptmx}
\usepackage{etoolbox}

\usepackage{color}
\usepackage{xcolor}

\usepackage{ulem} 

\newcommand{\dyadic}[1]{{#1}
\setbox0=\hbox{$\scriptstyle\leftrightarrow$}
   \setbox2=\hbox{$#1$}
   \dimen0=.5\wd0 \advance\dimen0 by-.5\wd2
   \advance\dimen0 by-\wd0
   \kern\dimen0
{^{\hbox{$\scriptstyle\leftrightarrow$}}}}

\makeatletter
\def\@email#1#2{%
 \endgroup
 \patchcmd{\titleblock@produce}
  {\frontmatter@RRAPformat}
  {\frontmatter@RRAPformat{\produce@RRAP{*#1\href{mailto:#2}{#2}}}\frontmatter@RRAPformat}
  {}{}
}%
\makeatother

\renewcommand{\selectlanguage}[1]{}

\begin{document}

\preprint{AIP/123-QED}

\title{Imaging of induced surface charge distribution effects in glass vapor cells used for Rydberg atom-based sensors}

\author{Link~Patrick}
\affiliation{Depart. of Phys., University of Colorado, Boulder,~CO~80305,~USA}

\author{Noah Schlossberger}
 \affiliation{National Institute of Standards and Technology, Boulder,~CO~80305, USA}
\author{Daniel F. Hammerland}\affiliation{Depart. of Phys., University of Colorado, Boulder,~CO~80305,~USA}
\author{Nikunjkumar~Prajapati}\affiliation{National Institute of Standards and Technology, Boulder,~CO~80305, USA}
\author{Tate~McDonald}\affiliation{Depart. of Phys., University of Colorado, Boulder,~CO~80305,~USA}
\author{Samuel~Berweger}\affiliation{National Institute of Standards and Technology, Boulder,~CO~80305, USA}
\author{Rajavardhan~Talashila}\affiliation{Depart. of Elect. Eng., University of Colorado, Boulder,~CO~80305,~USA}


\author{Alexandra~B.~Artusio-Glimpse}\affiliation{National Institute of Standards and Technology, Boulder,~CO~80305, USA}
\author{Christopher~L.~Holloway}
 \email{link@nist.gov, christopher.holloway@nist.gov}
 \affiliation{National Institute of Standards and Technology, Boulder,~CO~80305, USA}%

\date{\today}

\begin{abstract}
We demonstrate the imaging of localized surface electric (E) field effects on the atomic spectrum in a vapor cell used in Rydberg atom-based sensors. These surface E-fields can result from an induced electric charge distribution on the surface. Induced surface charge distributions can dramatically perturb the atomic spectrum, hence degrading the ability to perform electrometry. These effects become pronounced near the walls of the vapor cell, posing challenges for vapor cell miniaturization. Using a fluorescence imaging technique, we investigate the effects of surface charge on the atomic spectrum generated with electromagnetically induced transparency (EIT). Our results reveal that visible light (480 nm and 511 nm), i.e., the coupling laser used in two-photon Rydberg EIT schemes, generates localized patches of charge or dipoles where this light interacts with the glass walls of the vapor cell, while a three-photon Rydberg EIT scheme using only near-infrared wavelength lasers shows no measurable field induction. Additionally, imaging in a vacuum chamber where a glass plate is placed between large electrodes confirms that the induced charge is positive. We further validate these findings by studying the photoelectric effect with broadband light during EIT and impedance measurements. These results demonstrate the power of the fluorescence imaging technique to study localized E-field distributions in vapor cells and to target the photoelectric effect of the alkali-exposed glass of vapor cells as a major disruptor in Rydberg atom-based sensors.

\end{abstract}

\maketitle

\section{Introduction}

Rydberg atom-based electrometry is an exciting area of research moving towards real-world applications in sensing, imaging, communications, and thermometry\cite{Schlossberger2024Nature}. 
These systems rely on an alkali vapor filled glass cell, and one of the main hurdles for the field is developing vapor cells without stray electric (E) fields, which reduce detector sensitivity. Vapor cells developed for ground-state chip-scale magnetometers\cite{10.1063/1.5026238} and chip-scale atomic clocks (CSAC)\cite{10.1063/1.1787942} have been very successful over the past decade. However, the distinct sensitivities of CSACS and magnetometers, as well as variations in their production materials and methods, means that their developments are not distinctly transferable to Rydberg-atom based field sensors. Understanding the distributions and mechanism for producing stray fields in vapor cells is a necessity for improving sensitivity and robustness. 

In Rydberg electrometry, an applied radio frequency (RF) E-field perturbs the energy states of a Rydberg atom. The typical manner to readout the perturbed Rydberg state is with electromagnetically induced transparency (EIT).  EIT is a convenient way to measure not only the atomic state of a Rydberg atom, but also any perturbation in the state due to an applied RF E-field~\cite{Osterwalder_Merkt_1999}. When the RF field is on-resonance with an atomic transition, the EIT signal experiences Autler-Townes (AT) splitting~\cite{Schlossberger2024Nature, Sedlacek_Schwettmann_Kübler_Löw_Pfau_Shaffer_2012}. This AT splitting is directly related to the magnitude of the applied E-field and the atomic dipole moment of the particular atomic state, as such the AT approach leads to a direct International System of Units (SI)-traceable measurement of the RF E-field strength. When the applied RF field is non-resonant, the readout relies on Stark shifting of the atomic state and consequently a frequency shift of the EIT signal~\cite{Schlossberger2024Nature}, where the frequency shift is proportional to the square of the magnitude of the applied field and is proportional to the atomic polarizability. In order to obtain accurate and repeatable measurements when using either the AT approach or the Stark shift approach, a well-resolved EIT spectra is required.

\begin{figure}
    \centering
    \includegraphics[scale = 0.9]{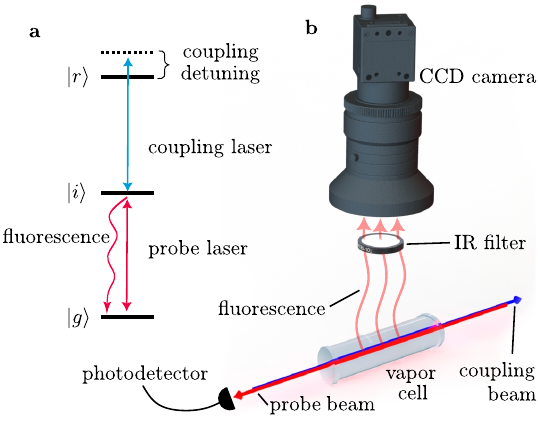}
    \caption{Experimental setup. A photo-detector is used detect the EIT signal and a camera is used to detector the fluorescence.}
    \label{fig:setup}
\end{figure}

When performing these EIT measurements in vapor cells, various factors can cause distortions to the EIT lineshapes, which manifest in frequency shifts and broadening of the measured EIT lineshape. In fact, there are many examples of distortion of the EIT signal in vapor cells, especially in the development of compact vapor cells (nominal dimensions of $<1cm$)\cite{Kohlhoff_2016}. There are several factors that cause the distortions of the EIT lineshapes. These include contaminant gases\cite{eric}, non-uniform E-fields\cite{10.1063/5.0161213}, background E-fields\cite{Ma_Paradis_Raithel_2020,PhysRevA.75.062903}, induced surface electric dipoles\cite{Sedlacek_Kim_Rittenhouse_Weck_Sadeghpour_Shaffer_2016}, and other systematic effects\cite{Carter_Martin_2011,Abel_Carr_Krohn_Adams_2011}. The manufacturing process can leave contaminant gases inside the cell\cite{eric}. Contaminant gases have the effect of broadening and shifting EIT lines. However, it has been shown that contaminant gases would have to be a fairly large fraction of the total pressure for this to be a problem\cite{eric}. As a result, it is most likely not the cause of poor vapor cell performance, where roughly 0.02~mbar of contaminant gas would add roughly 1~MHz of additional broadening. Non-uniform fields inside the vapor cell (due to internal RF resonances and/or non-uniform applied fields) can cause broadening and arbitrary EIT line shape\cite{10.1063/5.0161213}.  In the absence of an applied RF field, the EIT lineshape can exhibit distortions due to charges (or dipoles) on the surface of the vapor cell walls, which may be caused by photon-ionization\cite{Ma_Paradis_Raithel_2020}, Penning ionization, alkali atoms attaching to the surfaces creating electric dipoles~\cite{Sedlacek_Kim_Rittenhouse_Weck_Sadeghpour_Shaffer_2016}, surface contaminants\cite{PhysRevA.75.062903}, as well as other surface interactions.  These charges and dipoles cause DC Shark shifts and broadening to the EIT lines. However, these EIT measurements via the conventional photodetector methods do not contain information about the location or geometry of the charge distribution. Therefore, a spatially resolved measurement of the EIT along the beam is needed to better investigate the charge distribution inside vapor cells. 

In this paper, we report on a recently developed fluorescence imaging technique\cite{noahimaging} to investigate surface charge distributions and their resulting E-fields on EIT lineshapes. The EIT signal will show a Stark shift in the presence of an E-field produced by the surface charges. This fluorescence imaging approach allows us to determine spatially dependent the E-field strength from the surface charges along the length of the vapor cell. In this methodology, EIT is generated inside a vapor cell. The coupling beam is frequency swept over the EIT signal, yielding a fractional change in the fluorescence as the populations of states are altered. The fluorescence is imaged by a camera, as shown in Fig.~\ref{fig:setup} and a shift in the EIT lineshape indicates an E-field present at that location in 1-dimension. This approach allows us to observe localized E-fields along the laser propagation path. As such, we can investigate local charge effects in the laser propagation path.  The approach is advantageous over simply monitoring the EIT signal with a photodetector at the output of the probe light because the photodetector EIT signal is a path-integrated measurement of all the broadening and shifting mechanisms as the light propagates through the entire cell. As a result, isolated local effects are difficult to detect and locate in a simple photodetector EIT signal.

\section{Experimental Results}

We first performed fluorescence imaging using a standard $^{85}$Rb two-photon ladder EIT excitation scheme~\cite{Mohapatra_Jackson_Adams_2007}, where we used a 780~nm probe laser and a 480~nm coupling laser, see Fig.~\ref{fig:setup}. We generated an EIT signal for the 50D$_{5/2}$ Rydberg state. The probe and coupling powers incident on the cell were 76 $\mu$W and 380 mW respectively, and the full-width at half-maximum beam diameters were 590~$\mu$m and 1050 $\mu$m respectively. The fluorescence was collected through a 25~mm diameter filter at a distance of 15~cm (with the camera accepting light over a broader angular range), resulting in an effective numerical aperture of 0.08 and a collection efficiency of 0.2\%. The projected pixel size was calibrated by imaging a ruler in the beam plane. We used a cylindrical vapor cell of 78~mm length and 25~mm diameter.  In these experiments, we were able to obtain an imaging resolution of 50~$\mu$m.

\begin{figure*}
    \centering
    \includegraphics[scale = 0.9]{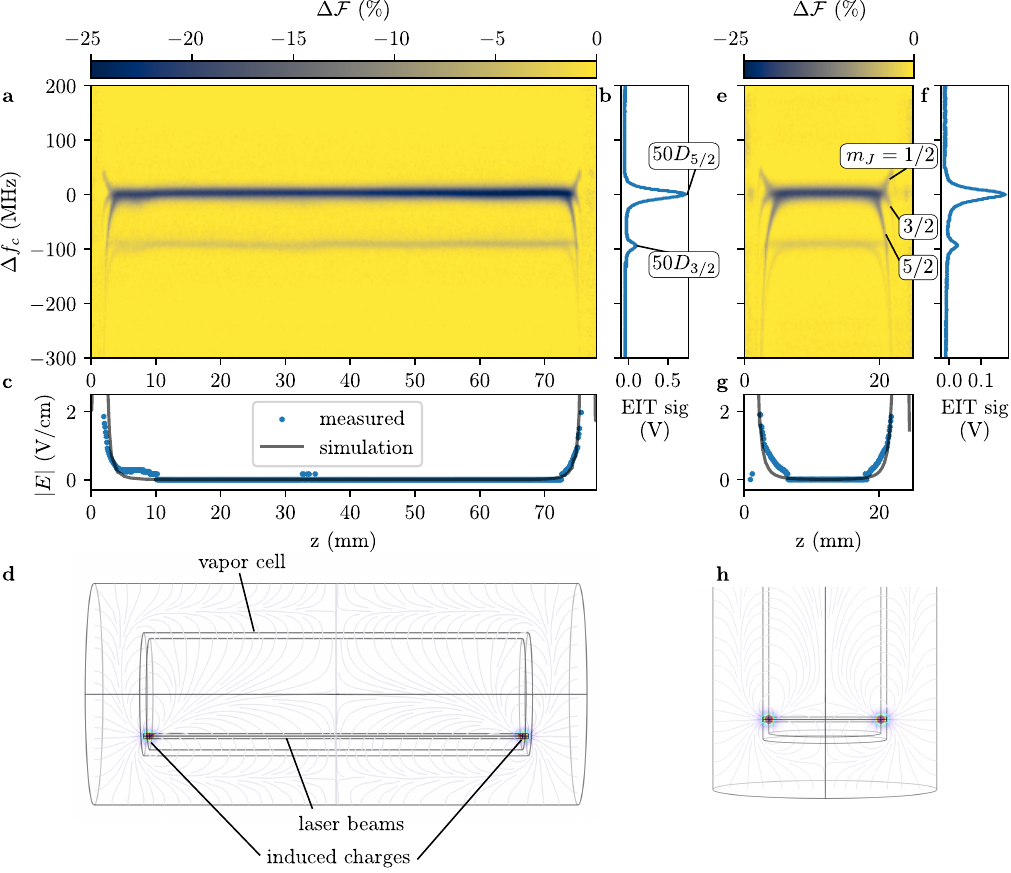}
    \caption{Images of spectra for the $5S_{1/2} \rightarrow 5P_{3/2} \rightarrow 50D_{5/2}$ ladder of $^{85}$Rb in a 25~mm diameter by 78~mm long cylindrical vapor cell. \textbf{a} Fluorescence spectrum as the coupling laser is scanned over the $50 D_{5/2}$ and $50 D_{3/2}$ states along the beam measured with the CCD camera. \textbf{b}. The corresponding EIT signal representing a change in transmission measured by the photodetector. \textbf{c} The fluorescence spectrum is fit to Eq. \ref{eq:ACStarkFit} to find the field at each point. This is compared to the field distribution predicted by the finite element model, shown in \textbf{d} with the predicted E-field strength and field lines. \textbf{e-h} The same is repeated when imaging along the short axis of the vapor cell.}
    \label{fig:bothpath}
\end{figure*}

Measurements were taken in two orientations of the vapor cell: with the beam propagating along the long axis (75~mm path length) and also along the short axis (25~mm path length) of the vapor cell. Fig.~\ref{fig:bothpath} shows both the EIT signal measured on a photodetector and the fluorescence imaging along the laser propagation path. We see more broadening of the EIT signal for the short propagation path than that for the long path. This is explained by observing the fluorescence image close to the laser entrance and exit locations of the vapor cell.  We see distinct peaks emerging from the different polarizabilities of the $m_j$ sub-levels in both Fig.~\ref{fig:bothpath} \textbf{a} and \textbf{e}. The observed spectral perturbations are due to the E-field caused by induced surface charges at the input and output locations of the lasers. 

\subsection{2-photon vapor cell measurements}

For small fields, the DC Stark shift of the EIT signal along the propagation axis is given by
\begin{equation}
    \Delta f_{\textrm{Stark}, m_J} \approx -\frac{1}{2}\alpha_{m_J} E^2/h,
\end{equation}
where $\alpha_{m_J}$ is the polarizability of a Rydberg state, $h$ is Planck's constant, and $E$ is the peak value of the E-field. For strong fields, on the order of ($\sim$V/cm), the response is no longer quadratic and a full Stark map must be used to calculate  $\Delta f_{\textrm{Stark}, m_J}$. Due to the varying polarizability of each $m_J$ sublevel, the field will cause the EIT signal to split into multiple distinct peaks. The change in fluorescence due to EIT in the presence of a stark shift is then

\begin{equation}
  \Delta \mathcal{F}_\mathrm{Stark} = \sum_{m_J} A_{m_J}e^{-\frac{(\Delta f_c - \Delta f_{\textnormal{Stark},m_J})^2}{2\sigma^2}}, \label{eq:ACStarkFit}
\end{equation}
where $\Delta f_c$ is the frequency detuning of the coupling laser, $\sigma$ is the spectral linewidth of the EIT feature, and $A_{m_J}$ are empirical weights related to relative polarizations of the optical and E-fields. We can resolve the field from the spectrum by fitting the spectrum to this equation. This is used to measure the E-field along the propagation path inside the vapor cell. 

Fig.~\ref{fig:bothpath} shows the measured E-field along the propagation path for each vapor cell orientation. We see that close to the cell walls, the estimated E-field levels are on the order of 1~V/m and we see that the field level decays as a function of distance from the wall.  This type of decay in distance of the E-field is indicative of a concentrated patch charge. 
We used a finite element model to numerically calculate the E-field from a patch charge as a function of distance from a surface. In our model, the inner walls of the vapor cell are grounded (as alkali atoms adsorb, the surface becomes more conductive \cite{PhysRevApplied.13.054034, PhysRevA.75.062903}), and a uniform surface charge density is placed on a 1~mm disc where the laser beams enter and exit the cell. These modeled results are compared to the experimental date in Fig.~\ref{fig:bothpath}(e) and (g), where we see good correlation. While the experiment is not sufficient to determine the exact charge distribution induced on the vapor cell, it is apparent that the fields we observe are only consistent with a highly localized distribution where the beams enter and exit the cell. In that, a decrease in an E-field level as a function of distance from the wall would not be possible if charges were uniformly distributed across a very large surface. For long propagation paths, the photodetector EIT is relatively unaffected as the majority of the signal comes from portions of the vapor cell which do not experience these effects. However, this poses a challenge for the miniaturization of vapor cells with short propagation path, as the Stark shifted atoms will comprise the entire EIT signal for cells smaller than \textasciitilde1~cm.

\begin{figure}
    \centering
    \includegraphics[scale = 0.9]{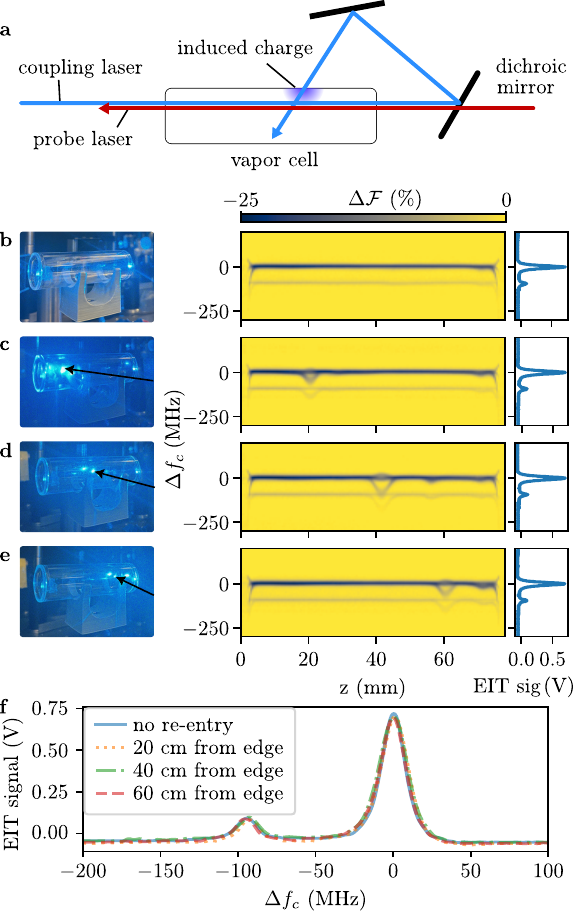}
    \caption{Re-entry of the visible light to the side wall of the vapor cell using the $5S_{1/2} \rightarrow 5P_{3/2} \rightarrow 50D_{5/2}$ ladder of $^{85}$Rb. \textbf{a} A diagram of the setup. \textbf{b} The cell (imaged on the left) is measured via fluorescence imaging (middle) and transmission (right) without the re-introduction of the visible beam. \textbf{c} The beam is re-introduced 20~mm from the left of the cell, \textbf{d} 40~mm from the left of the cell, and \textbf{e} 60~mm from the left of the cell. \textbf{f} The EIT signals in the four cases are overlayed.}
    \label{fig:reentry-setup}
\end{figure}

These results imply that the light (likely only the visible light, 480~nm as shown below) causes induced charge distribution on the glass surfaces at the input and output locations of the lasers. In order to verify this, we perform a set of experiments where the visible light (480~nm laser) is recycled into the vapor cell in a direction that is orthogonal to the EIT measurement path, see Fig.~\ref{fig:reentry-setup}. With the fluorescence imaging, we are able to investigate the re-entry location of the visible light. Fig.~\ref{fig:reentry-setup} shows the fluorescence and the estimated E-field along the propagation path for different re-entry point (along the z-axis). In these results, the probe and coupling lasers are placed 1~mm from the side wall and are propagating along the z-axis. Without the re-entry of the coupling laser, we see no Stark shift (i.e., no induced charges) along the of propagation path $z$ (except at the ends of the vapor cell: input and output location of the EIT lasers). When the visible light is re-entered, we see notable Stark shifts at the reentry locations. This indicates that the visible light is inducing surface charges. The EIT signals measured via the photodetector are very similar regardless of the location of the reentry point along the cell wall. As a result, a single photodetector EIT measurement cannot determine the location of the surface charge.

\begin{figure}
    \centering
    \includegraphics[scale = 0.9]{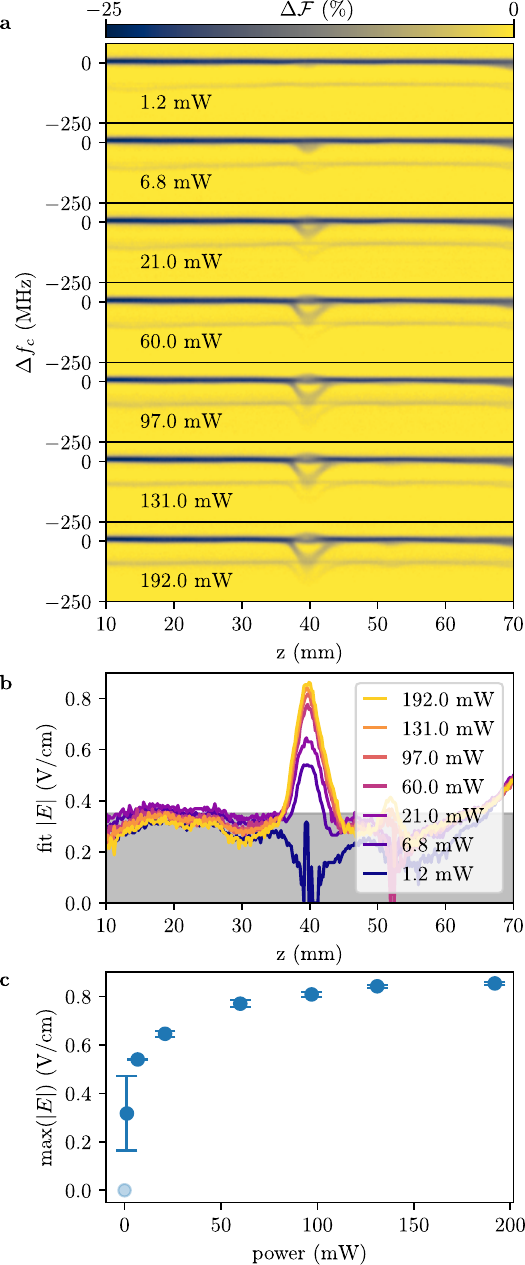}
    \caption{\textbf{a} Fluorescence measurements of the EIT as a function of optical power incident on the glass as the coupling laser is swept over the 50D$_{5/2}$ and 50D$_{3/2}$ states, indicating an increase in amount of charge with higher optical power. \textbf{b} The fluorescence spectrum is fit to the Stark map to determine the E-field magnitude at each location along the vapor cell. The threshold at which the field cannot be resolved is greyed out. \textbf{c} The maximum E-field magnitude at the point of coupling light re-entry is plotted as function of the optical power.}
    \label{fig:powersweep}
\end{figure}

The induced E-field strength as a function of visible optical power is investigated in Fig.~\ref{fig:powersweep}. With the re-entry of the coupling laser at a position near the middle of the cell, the optical power is varied from 1.2-192~mW. We observe that stronger E-fields and therefore more charge is generated with higher incident optical power. The results in Fig.~\ref{fig:powersweep}(e) depict a saturation behavior with increasing power. Due to the unknown charge distribution, the actual amount of charge generated can not be easily determined. 

\subsection{Positive charge measured with electrodes in a chamber}

\begin{figure}
    \centering
    \includegraphics[width=0.9\linewidth]{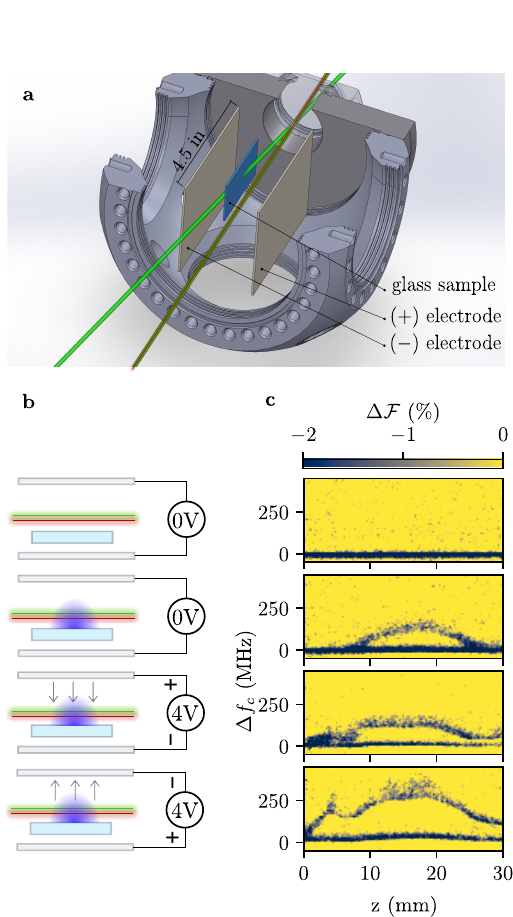}
    \caption{\textbf{a} Vacuum chamber that is equipped with plate electrodes and a fixture to hold glass samples. \textbf{b} Diagrams of the setups corresponding to the fluorescence images. \textbf{c} Fluorescence images from Cs 42D$_{3/2}$ in a vacuum chamber with visible light (510~nm) striking a glass sample positioned 3~mm from the Rydberg beam: no visible light, visible light with no voltage, visible light with $+4$~V, and visible light with $-4$~V.}
    \label{fig:chamberwithVoltage}
\end{figure}

It is instructive to ascertain the direction of the E-field produced at the surface. To determine this, we performed a set of experiments within a vacuum chamber that is equipped with plate electrodes and a fixture to hold glass samples, see Fig.~\ref{fig:chamberwithVoltage}.  The vacuum chamber contains $^{133}$Cs atoms. Inside the chamber we generate a two-photon ladder $^{133}$Cs EIT signal using the 6S$_{1/2}$-6P$_{3/2}$-42D$_{3/2}$ excitation path., which requires a 850~nm probe and a 511~nm coupling laser. Once the EIT signal is generated, we can move the glass plate to a position such that the EIT lasers are propagating along the glass surface without touching the glass (3~mm away). We recycled the visible light (511~nm in this case) after it had propagated through the chamber and directed it back onto the glass plate. The fluorescence imaging is shown in Fig.~\ref{fig:chamberwithVoltage}, and indicates that the visible 511~nm light induces a surface charge distribution.

We then applied $\pm4$~V to the plate electrodes. The plus and minus voltage on the plate introduces an E-field that enhanced or diminished the E-field caused by the surface charge distribution. We see in Fig.~\ref{fig:chamberwithVoltage} that for a negative voltage the E-field is enhanced and for a positive voltage the E-field is repressed. These results indicate that the induced surface charge, at the location of incident light, is positive.

\subsection{Validation via broadband light source}

\begin{figure}
    \centering
    \includegraphics[width=0.9\linewidth]{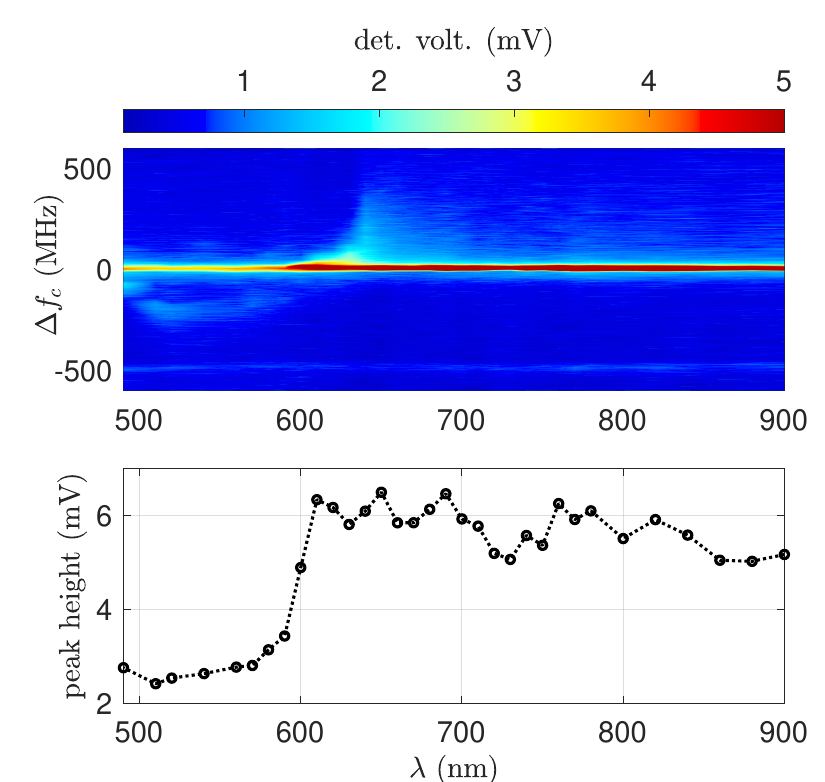}
    \caption{EIT measured via photodetector as a function of incident wavelength from a broadband light source indicating that the produced charge is due to the photoelectric effect. }
    \label{fig:photoelectric_super}
\end{figure}

To better understand the wavelength sensitivity of the induced charges, a broadband laser source is used to irradiate the vapor cell to observe changes in the EIT.
A super continuum source generates a broadband optical spectrum from 490-900 nm. The position of a fixed width slit in the Fourier plane allows for the output wavelength to be tuned. 
The vapor cell was illuminated with 0.2-2~mW of optical power by the broadband source for the wavelengths 490-900~nm, respectively. A single photodetector was used to monitor the EIT signal via the 2-photon detection scheme ($^{133}$Cs | 6S$_{1/2}$-6P$_{3/2}$-52D$_{5/2}$). As shown in Fig.  ~\ref{fig:photoelectric_super}, the EIT signal indicates the presence of an E-field only with incident light below 600~nm (2.1~eV). This aligns well with the literature work function of $^{133}$Cs, around 1.9-2.1~eV ~\cite{Kawano_2022,Aghili_Rahbarpour_Berahman_Horri_2024, Nikolić_Radić_Minić_Ristić_1996}. As a result, we suspect that the $^{133}$Cs on the surface of glass is ionized by any visible light with a wavelength less than 600~nm. 

To verify that the photoionization originates from the condensed $^{133}$Cs and not the glass, we measured the conductivity of a vapor cell and a piece of borosilicate glass (the same glass of the vapor cell) as a function of incident wavelength. The impedance of the vapor cell will change as charges are generated in between the electrodes used to make the impedance measurement\cite{Bouchiat_Guéna_Jacquier_Lintz_Papoyan_1999}. The setup for the impedance measurement with the glass sample is shown in Fig.~\ref{fig:conductElectric}.  As shown, the measured impedance is affected by the photoelectric effect around the same wavelength as the EIT signal, and the impedance of just a piece of glass does not show any response to the incident light. These results show that unlike the vapor cell with $^{133}$Cs, we do not observe a wavelength dependence on the impedance for the glass sample.

\definecolor{myYellow}{rgb}{1, 0.7, 0.1} 

\begin{figure}
    \centering
    \includegraphics[width=0.85\linewidth]{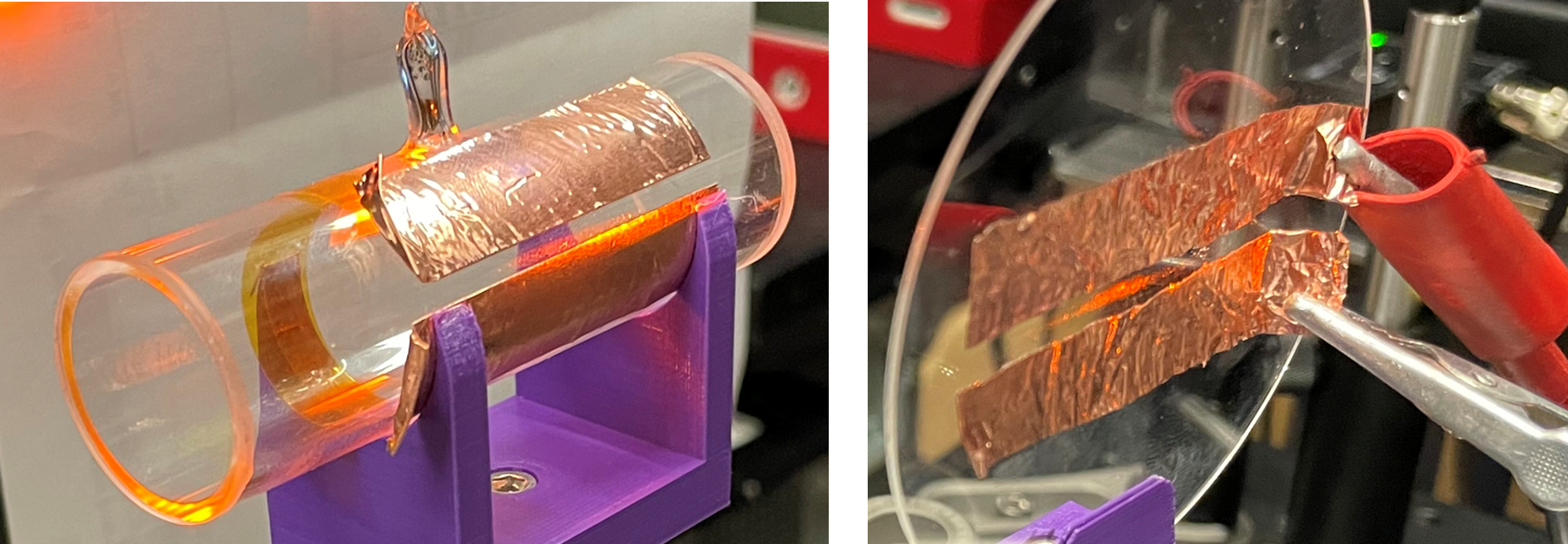}
    \includegraphics[width=0.9\linewidth]{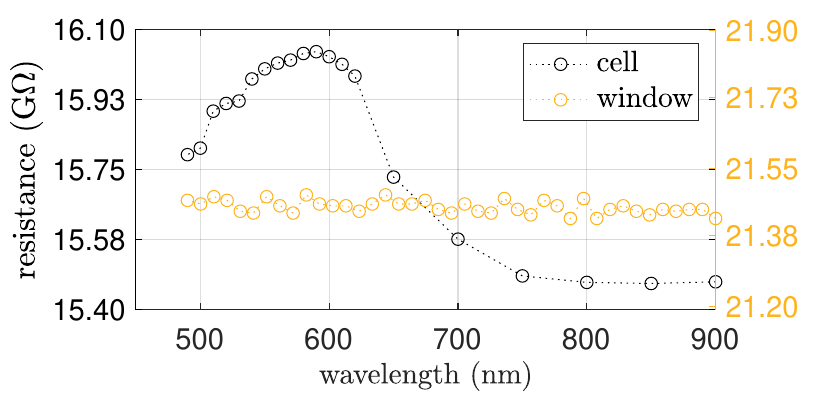}
    \caption{(top) Impedance measurement setup with electrodes adhered to the surface of the glass. The orange 600~nm light from the broadband source is shown illuminating the samples: vapor cell (left) and window (right). Resistance of a Cs vapor cell as a function of incident light (black). Resistance of a borosilicate glass as a function of incident light (\textcolor{myYellow}{yellow}).}
    \label{fig:conductElectric}
\end{figure}

\subsection{Validation via 3-photon scheme}

\begin{figure*}
    \centering
    \includegraphics[scale=0.9]{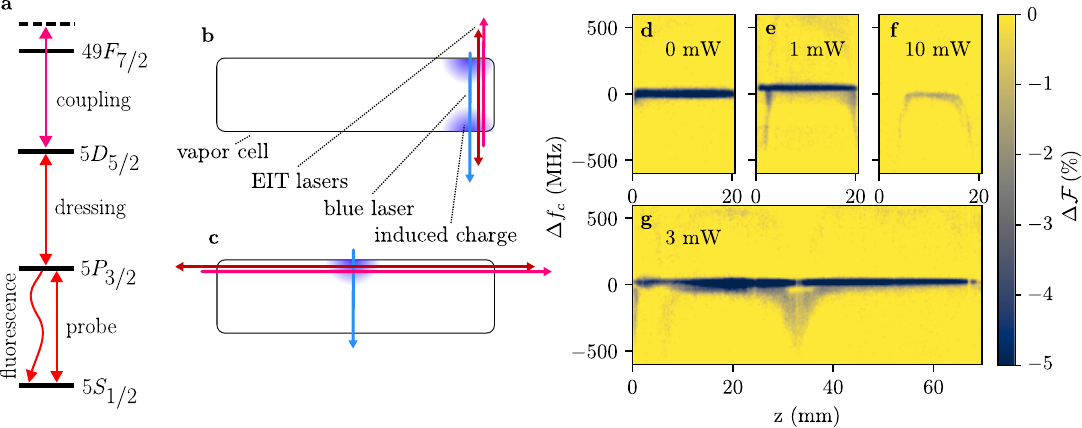}
    \caption{\textbf{a} Level diagram for three-photon system. \textbf{b} Diagram of the co-linear setup. \textbf{c} Diagram of the orthogonal setup. \textbf{d} Fluorescence along cell length as coupling laser frequency is scanned across the resonance with no blue laser present. \textbf{e} 1 mW of blue laser power is added co-linearly with the other laser beams. \textbf{f} 10 mW of blue laser power is added co-linearly. \textbf{g} 3 mW of blue laser power is added orthogonally incident on the side of the cell.}
    \label{fig:three-photon_face_entry}
\end{figure*}

To further validate that the visible light (480~nm) is the main culprit in inducing surface charges, we performed a set of experiments with a three-photon ladder EIT excitation scheme (which requires no visible light source) in $^{85}$Rb. This 3-photon scheme uses the 5S$_{1/2}$-5P$_{3/2}$-5D$_{5/2}$-49F$_{7/2}$ excitation path, which requires a 780~nm, 776~nm, and 1259~nm laser, shown by the level diagram in Fig.~\ref{fig:three-photon_face_entry} (a).  We use the same 75~mm by 25~mm vapor cell as before. 

The three EIT lasers were positioned $2$~mm from the parallel cell wall as illustrated in Fig.~\ref{fig:three-photon_face_entry} (b). We measured fluorescence imaging for this orientation along the 25~mm path. Unlike the results for the two-photon scheme (using visible light), we see little-to-no effect at the locations where the lasers enter and exit the vapor cell, shown in Fig.~\ref{fig:three-photon_face_entry} (d). We then introduced a blue 480~nm laser (off resonant to any atomic transition) co-linearly along the propagation path with the three EIT lasers. We repeated the measurements with blue laser optical powers of 1~mW and 10~mW, shown in Fig.~\ref{fig:three-photon_face_entry} (e and f). We observe the expected Stark shift due to induced charges on the glass surface from the blue laser. Then the vapor cell is rotated and the blue laser is redirected to enter the vapor cell orthogonal to the other laser paths as shown in Fig.~\ref{fig:three-photon_face_entry} (c). We set the re-entry location of the coupling 480~nm laser to $z=38$ mm. When the 480~nm laser is on, large Stark shifts are again observed and the location of where the 480~nm light enters the cell as shown in Fig.~\ref{fig:three-photon_face_entry} (g). We can see this effect for a laser power as low as 1 mW. We observed a shift of nearly 200~MHz for a power as low as 3~mW. We also see a response when using a blue laser with powers as low as 0.5~mW. This illustrates the sensitivity of a highly polarizable Rydberg state (the 49F$_{7/2}$ state used here) and the detrimental impact of induced charges on precise measurements of weak E-fields.

\section{Discussion}

These results indicate that the optical beams interacting with the alkali-vapor and glass interface  are the dominate source of stray fields in the vapor cells. It is likely, given the complexity of the alkali-glass interface, that there are other mechanisms that lead to the charge distributions that are reported in this article. The most simplistic and dominant source of charge distributions is a direct photoionization of the thin, metallic layer of Cs or Rb that innately condenses on the glass surface in saturated vapor cells. The 1.9 to 2.1 eV binding energy of bulk Cs means that light with a wavelength of less than 650 nm will successfully ionize the metal. This boundary is well produced in the resistance measurements shown in Fig.~\ref{fig:conductElectric}, where the resistance is unchanged until approximately 650 nm \cite{Nikolić_Radić_Minić_Ristić_1996, Kawano_2022}. The slightly earlier onset of the effect near 700 nm likely arises from the bandwidth of the source, which we measured to be 20~nm while set to emit 700~nm light. Furthermore, it has been shown that the exact geometry of the metal surface \cite{Aghili_Rahbarpour_Berahman_Horri_2024} and exposure to different materials \cite{Wells_Fort_1972} can shift the ionization energies. The ionization energy of bulk alkali also explains the difference in charging observed in the two and three photon experiments. Finally, the vacuum chamber electrode experiments point towards a positive charge on the surface of the glass – these are likely the residual positive Cs ions after the photoionization process. From these results, we conclude that a direct photoionization of bulk alkali on the vapor cell wall is a primary contributor to the stray fields.

Other mechanisms may still be producing stray fields/charges inside the vapor cell, albeit to a lesser degree than the photoelectric effect, and may be the subject of future research. The chemistry of alkali metals with glass is another likely suspect of the surface charges. It has been shown that non-bridging oxygen defects in SiO$_2$ lattices react strongly with cesium atoms. Under this interaction, the valence electron in cesium is almost completely consumed by the surrounding oxygen atoms and an Cs+ ion is effectively bound onto the surface \cite{Lopez_Illas_Pacchioni_1999}. This dipole-like charge distribution would exist independently of the wavelengths incident on the vapor cell. While no EIT can be measured on the pure glass itself, the impedance measurements taken on the borosilicate wafer indicate that the glass itself is not a major contributor to the effect for wavelengths larger than 480~nm. However, stray UV light may cause charges from the glass itself and should still be taken into consideration for sensitive measurements.

\section{Conclusion}

We demonstrate a measurement method for imaging and investigating the location of surface charge effects on the inner walls of vapor cells. We show that visible light (480~nm and 511~nm) induces a charge distribution on the surface of vapor cells at the entry/exit point locations of the visible light.  We show that by using a three-photon EIT scheme (no visible light sources), the effects are substantially reduced. This study illustrates the importance of developing manufacturing strategies for the surface of vapor cells in order to mitigate  the formation of surface charges when using two-photon excitation schemes, which involves visible light sources (480~nm for Rb and 511~nn for Cs). These studies also indicate that the three-photon EIT scheme may be preferred over two-photon EIT schemes if one is interested in developing compact vapor cells for Rydberg electrometry.

Fluorescence imaging is used to investigate the induced charge distribution inside an alkali filled vapor cell. The distortion of the EIT lineshape as a result of the induced charge is shown for both 2-photon and 3-photon excitation schemes. Fluorescence imaging of the EIT allows for a direct measurement of the localization of the induced charge distribution. EIT signal measurements with a simple photodector cannot be used to determine the location of the surface effect. The E-field strength from the induced charge as a function of optical power is measured using the fluorescent imaging technique. Electrodes within an alkali containing vacuum chamber are used to determine that the induced charge is positive at the point of visible light incidence. To verify that this is due to the photoelectric effect the EIT spectrum and also the impedance of the vapor cell is measured as a function of incident light wavelength. 

We are carrying out further investigations to better understand and quantify how charges or surface dipoles are induced by the visible light sources, as well as develop imaging techniques that will allow us to distinguish the difference between charge and dipole effects. Enhancement techniques such as lock-in detection will be used to improve the sensitivity for future measurements, as well as polarization control to provide the directionality of the electric fields. Also, our vacuum chamber in conjunction with the fluorescence imaging approach, gives a needed diagnostic tool for investigating various coating and other surface effects on glass in the presence of alkali metals. This will be the topic of a further publication.

\section*{Acknowledgments}
This work was supported by DARPA under the SAVaNT program and by the National Institute of Standards and Technology (NIST) through the NIST-on-a-Chip
(NOAC) program.  The views, opinions, and/or findings expressed are those of the authors and should not be interpreted as representing the official views or policies of the Department of Defense or the U.S. Government. A contribution of the U.S. government, this work is not subject to copyright in the United States. Distribution Statement "A" (Approved for Public Release, Distribution Unlimited).

\subsection*{Conflict of Interest}
The authors have no conflicts to disclose.
\vspace{3mm}
\subsection*{Data Availability Statement}
The data related to the findings of this paper are publicly available at doi.org/10.18434/mds2-3685.

\bibliography{chargebib}

\end{document}